\documentclass[fleqn,usenatbib]{mnras}
\usepackage{newtxtext,newtxmath}
\usepackage[T1]{fontenc}
\usepackage{graphicx}
\usepackage{amsmath}
\usepackage{booktabs}
\hypersetup{hypertexnames=false}
\DeclareRobustCommand{\VAN}[3]{#2}
\let\VANthebibliography\thebibliography
\def\thebibliography{\DeclareRobustCommand{\VAN}[3]{##3}\VANthebibliography}
\newcommand{\source}{Gaia DR3 2981818948336660224}
\title[An unresolved warm infrared excess]{An unresolved extreme warm infrared excess\\toward a nearby low-mass star}
\author[S. Torlakcik]{Sahin Torlakcik$^{1}$\\
$^{1}$Ankara Atat\"urk High School, Ankara, T\"urkiye}
\date{Preprint, September 2026}
\pubyear{2026}
\begin{document}
\label{firstpage}
\pagerange{\pageref{firstpage}--\pageref{lastpage}}
\maketitle
\begin{abstract}
Gaia DR3 2981818948336660224 is a nearby low-mass star with a substantial warm mid-infrared excess. Identified in a Gaia--2MASS--WISE search, it lies at approximately 127 pc and has an AllWISE W3 signal-to-noise ratio of 33.1. We recover the emission in all 16 available cryogenic exposures over 1.26 d; removing any one exposure leaves the signal intact. The W3 emission is centred close to the propagated stellar position, and the known neighbour 6.1 arcsec away fails to reproduce its morphology. Nominal scene fits place 97--98 per cent of the W3 flux at the target position. This provides strong spatial support for association, although template uncertainty and a nearly coincident hidden source prevent a definitive decomposition. The excess can be described by a roughly 300--350 K component with $L_{\rm IR}/L_\star\simeq0.022$, assuming stellar association and retaining the weaker W4 detection. If produced by circumstellar dust, it falls in the extreme warm-debris regime around a dwarf with no established evidence of youth. Its origin remains unresolved. The repeatable excess and close spatial association make this a useful target for subarcsecond MIR imaging and spectroscopy to test the warm-dust interpretation.
\end{abstract}

\section{Introduction}
Large amounts of warm dust are difficult to sustain around a main-sequence star. Extreme debris systems, with infrared fractional luminosities $f_{\rm IR}=L_{\rm IR}/L_\star\gtrsim0.01$, are therefore useful places to look for recent dust production. Their changing infrared emission can constrain how material is produced and removed \citep{KennedyWyatt2013,Moor2021,Moor2024}. Youth is part of this picture, but not the whole of it: recent censuses contain many systems older than 100 Myr.

The low-mass hosts deserve particular attention. Warm excesses are uncommon in WISE studies of nearby field M dwarfs, while disc-bearing stars are found in young M-dwarf samples \citep{Avenhaus2012,BinksJeffries2017}. These surveys have different selections and sensitivities. Even so, a percent-level warm component around a low-mass dwarf without established signs of youth is worth a closer look.

The difficulty is deciding where the infrared light comes from. A background galaxy can sit inside a WISE beam and produce an apparently stellar excess \citep{Ren2024,Blain2024}. Radio imaging of Hephaistos candidate G identified an AGN near the star \citep{Ren2025}; subsequent archival diagnostics examined this problem across the candidate sample \citep{Ren2026}. Recent JWST observations resolved this problem for two Project Hephaistos sources, locating the excess in galaxies about an arcsecond from the stars \citep{Zackrisson2026}. This matters both for debris-disc studies and for thermal technosignature searches \citep{Wright2014,Suazo2024}.

Here we present a source-level investigation of \source. Its W3 excess survives remeasurement of the individual exposures and tests of the known neighbour. The emission is concentrated near the Gaia position. The repeated detections and the failure of the neighbour-only model narrow the problem: the remaining question is whether the central MIR source belongs to the star or to a much closer, unresolved object.

\section{Source and discovery}
\begin{figure}
\centering
\includegraphics[width=\columnwidth]{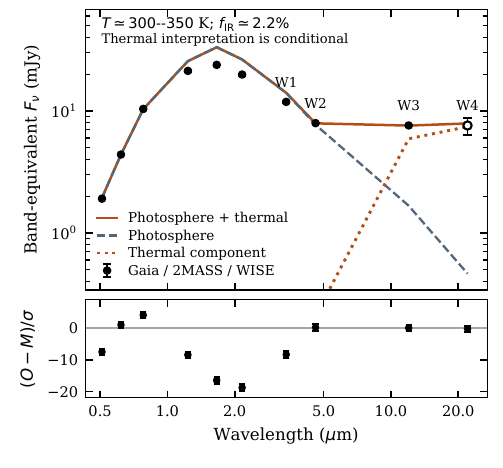}
\caption{Optical-to-MIR SED with the stellar and thermal components. W3 and W4 lie above the photosphere, and a warm component follows the MIR measurements. Points and model samples are band-integrated signals shown as equivalent flux densities; the connecting lines are guides. Residuals in units of formal measurement error expose the remaining optical/NIR mismatch, discussed in Section~4.}
\label{fig:sed}
\end{figure}

The source was identified in Dyson Atlas, a multimodal open-source infrared-excess technosignature survey of 20.8 million Gaia--2MASS--WISE stars. Papers describing the survey methods and the full search are in preparation. Other searches in this field include Project Hephaistos \citep{Suazo2022,Suazo2024} and the data-driven search for W1/W2 excesses among FGK stars by \citet{Contardo2024}. The unusual excess of the present source motivated a dedicated investigation ahead of the broader Dyson Atlas publications. Table~\ref{tab:source} lists the adopted stellar properties, and Fig.~\ref{fig:sed} shows the spectral energy distribution (SED).

Gaia DR3 measures a parallax of $7.8814\pm0.0196$ mas and total proper motion of $13.59$ mas yr$^{-1}$ \citep{GaiaDR3}. The inverse parallax gives 126.88 pc before a zero-point correction, close to the GSP-Phot estimate of 126.55 pc; we use approximately 127 pc throughout. GSP-Phot gives $T_{\rm eff}=3768$ K and $\log g=4.73$, while the FLAME mass, radius and luminosity are consistent with a compact low-mass main-sequence dwarf \citep{Creevey2023}. FLAME formally favours a multi-Gyr age, but the slow evolution of a roughly $0.54\,M_\odot$ dwarf makes that age poorly constrained. The available solution does not indicate a young pre-main-sequence object, although independent youth diagnostics are still needed.

The kinematics strongly favour a field-star classification. Using Gaia astrometry and radial velocity, BANYAN $\Sigma$ \citep{Gagne2018}, with the updated population models of \citet{Gagne2026}, assigns a field probability of 99.6 per cent, or 98.4 per cent when radial velocity is omitted. We also find no entry for the source in the SPYGLASS-IV young-star membership catalogue \citep{Kerr2023}. These results argue against membership in a known young association, although they do not measure the stellar age or exclude a young, dispersed field star.

There is also a useful piece of source history. The position appears as PSO J078.89026$-$17.91014 in the variable-quasar candidate catalogue of \citet{Usatov2018}. That selection was photometric, without spectroscopic confirmation. Gaia's parallax and proper motion now identify the dominant optical counterpart as a nearby star; the possible presence of an additional MIR source is a separate question.

\begin{table}
\caption{Adopted source properties. Gaia coordinates are ICRS at J2016.0; stellar parameters are rounded catalogue model estimates. WISE magnitudes are on the Vega system with catalogue errors. Thermal values assume association with the star, and the temperature range spans the two fitting approaches described in Section~4 rather than a confidence interval.}
\label{tab:source}
\centering
\begin{tabular}{ll}
\toprule
Quantity & Value \\
\midrule
Gaia DR3 identifier & 2981818948336660224 \\
RA, Dec (deg) & $78.89026696$, $-17.91013761$ \\
Parallax (mas) & $7.8814\pm0.0196$ \\
Distance (pc) & $\simeq127$ \\
$T_{\rm eff}$ (K), $\log g$ (cgs) & $3768$, $4.73$ \\
$M_\star$, $R_\star$ (solar units) & $0.541$, $0.527$ \\
$L_\star/L_\odot$ & $0.0505$ \\
RUWE & $0.9935$ \\
W1 (mag) & $11.042\pm0.024$ \\
W2 (mag) & $10.838\pm0.021$ \\
W3 (mag) & $9.050\pm0.033$ \\
W4 (mag) & $7.604\pm0.172$ \\
W3, W4 catalogue S/N & $33.1$, $6.3$ \\
Thermal temperature (K) & $\simeq300$--$350$ \\
$L_{\rm IR}/L_{\rm direct\,\star}$ & $\simeq0.022$ \\
Known-neighbour separation & $\simeq6.1$ arcsec \\
\bottomrule
\end{tabular}
\end{table}

\section{W3 validation and source association}
\begin{figure*}
\centering
\includegraphics[width=\textwidth]{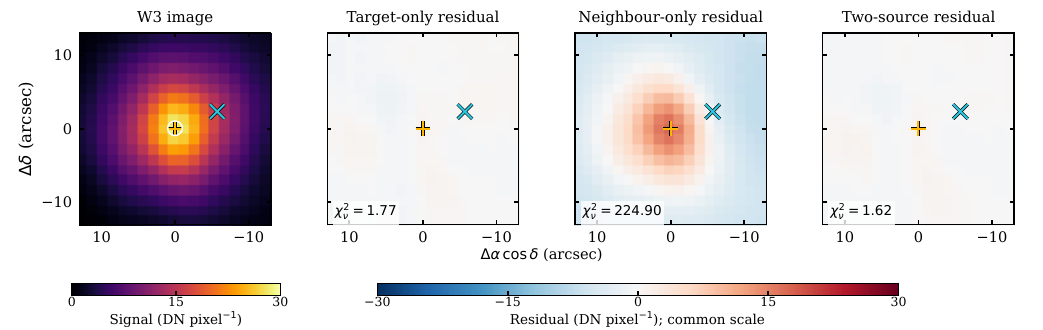}
\caption{W3 image and residuals from the tested source models. The orange plus marks the propagated Gaia position, the cyan cross the 6.1-arcsec neighbour and the white circle the measured centroid. Placing the source only at the neighbour leaves a strong central residual; models containing a source at the target position perform much better. All residual panels share a scale. The fits use a pole-coadd PSF and diagonal pixel errors, so their statistics are diagnostic. A source exactly coincident with Gaia would have the same image template.}
\label{fig:association}
\end{figure*}

\begin{figure}
\centering
\includegraphics[width=\columnwidth]{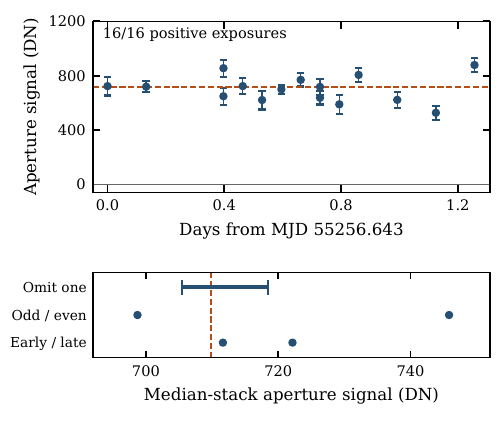}
\caption{W3 recovery in the 16 cryogenic exposures. Upper panel: aperture fluxes with the empirical blank-aperture scatter. Lower panel: subset stacks and the leave-one-out minimum-to-maximum range, which is not an uncertainty interval. Every exposure is positive, and omitting any one of them leaves the stacked signal intact. Fluxes are instrumental aperture sums in DN.}
\label{fig:repeatability}
\end{figure}

The catalogue evidence is encouraging. AllWISE\footnote{\url{https://wise2.ipac.caltech.edu/docs/release/allwise/expsup/}} reports W3 S/N $=33.1$, quality A, contamination flags \texttt{0000}, no saturation and a profile-fit reduced $\chi^2=1.084$ \citep{Wright2010}. Its deblending entries, $\mathtt{nb}=2$ and $\mathtt{na}=1$, make an image-level check necessary. Figure~\ref{fig:association} shows the W3 image and scene models; Fig.~\ref{fig:repeatability} summarises the individual exposures.

We remeasure all 16 available footprint-selected cryogenic W3 images at the propagated Gaia position. Each measurement uses a 3.25-arcsec-radius aperture and a clipped background plane fitted over a 19.5--32.5 arcsec annulus, with known neighbours masked. All 16 apertures have positive flux. Their flux-to-blank-aperture-scatter ratios exceed 5, with a median of 12.58. These are empirical noise diagnostics, rather than Gaussian detection significances.

No single exposure drives the result. The median frame flux is 718.50 DN, with a scaled median absolute deviation of 124.60 DN. The aligned median stack gives 709.85 DN, and the leave-one-out stacks span just 705.42--718.42 DN. Odd/even, early/late and worst-frame removals preserve the signal as well. These are instrumental aperture sums, including any emission within the aperture, not calibrated total-source fluxes. The 16 local footprint-selected images differ from the catalogue's 14 selected detections. They cover 1.257 d and share WISE systematics, so the result establishes repeatability over that interval rather than persistence over years.

The spatial evidence also favours the target position. After background subtraction, the positive-weight W3 centroid is offset by 0.065 arcsec. This small offset is descriptive; it is not a calibrated astrometric precision. The nearest known source, Gaia DR3 2981819017056137088, lies 6.1 arcsec away and has parallax $2.2715\pm0.0276$ mas, placing it well behind the target rather than in a physical wide pair.

We fit non-negative source amplitudes on identical pixels using the published W3 pole-coadd PSF and constant or planar backgrounds. Target-only, neighbour-only and two-source models give $\chi^2=498.0$, 63421.4 and 455.5, respectively. The neighbour-only model fails to reproduce the central peak. This is a strong geometric constraint, even though PSF mismatch and correlated pixels prevent a probability interpretation of the fit differences.

The nominal two-source fit assigns 97.8 per cent of the flux to the target position. Including a background plane and the 17.4-arcsec catalogue source changes this to 96.9 per cent. That agreement supports a target-dominated scene among the known sources. The exact fraction is less secure: changes in template width, rotation, position, fit radius and background produce a range of 57.3--100 per cent. This is a sensitivity envelope, not a confidence interval or a bound on hidden-source contamination. The PSF lacks matched field-control validation, and full pixel covariance is unavailable. Older centroid and decomposition estimates used different apertures, backgrounds and Gaussian templates, accounting for their different results.

A hidden source must satisfy a tighter positional requirement than the known neighbour. A grid covering $\pm2$ arcsec in 0.1-arcsec steps, with three PSF widths, places the preferred hidden-source positions 0.1--0.224 arcsec from Gaia. The contours are model-dependent and are not calibrated confidence regions. At exact coincidence, the hidden-source and stellar-position templates become identical. WISE can then locate the central emission but cannot divide its flux between the two objects. This is the remaining association problem.

W4 carries less weight. Its catalogue S/N is 6.3, the median frame diagnostic ratio is 1.86, and only two of 16 frames exceed 3; none exceeds 5. We retain it as a weak measured detection in the SED. Astrometry introduces another caveat: AllWISE's fitted motion, $(+244\pm36,+81\pm34)$ mas yr$^{-1}$, disagrees with Gaia despite the close propagated positional match. Later NEOWISE W1/W2 data do not resolve this inconsistency for W3 or provide a W3 motion measurement.

\section{SED and physical interpretation}
The excess is large compared with the expected photosphere. Using Gaia $G,G_{\rm BP},G_{\rm RP}$, 2MASS $J,H,K_{\rm s}$ \citep{Skrutskie2006} and WISE W1--W4, the baseline Gaia-parameter atmosphere predicts only about one-fifth of the observed W3 signal: the observed-to-predicted ratio is 4.65. The fits integrate a BOSZ 2024 atmosphere \citep{Meszaros2024} and a thermal component through the passbands, including foreground extinction. Broader stellar-parameter and extinction stress tests retain a positive W3 residual.

The source-level fit gives $T_{\rm IR}=301.7$ K and $L_{\rm IR}/L_{\rm direct\,\star}=0.02198$. The earlier frozen search fit gives 346.4 K. This difference comes from the fitting procedure, not new photometry: the search uses a discrete temperature grid, a fixed optical/NIR stellar normalisation, band-dependent discrepancy corrections and correlated prediction errors; the source fit interpolates temperature and allows a common stellar scale to vary across ten bands with formal diagonal errors. The search fraction of 0.02289 is defined relative to direct plus reradiated luminosity, equivalent to 0.02343 relative to direct starlight. A useful summary is therefore roughly 300--350 K and 2.2 per cent, rather than a precisely measured temperature and fractional luminosity.

The MIR shape is easier to describe than the full SED. With formal catalogue errors, the fixed-star thermal fit has $\chi^2=840.0$ for seven nominal degrees of freedom. Allowing seven stellar, extinction and thermal parameters reduces this to 72.9 for three, still a poor fit. Optical/NIR residuals, atmosphere errors and cross-survey calibration remain unresolved. Weak W4 and uncertain flux ownership add further uncertainty to the thermal parameters, so we do not quote formal intervals as their full errors.

If the emission is circumstellar dust, $f_{\rm IR}\simeq0.022$ places it in the extreme-debris regime. The comparison sample includes BD+20 307, ID8 and HD 23514 \citep{Song2005,Meng2015}, TYC 8830-410-1 and V488 Per \citep{Melis2021,Rieke2021}, and RZ Psc \citep{deWit2013}. TYC 8241-2652-1 illustrates how quickly a large warm excess can disappear \citep{Melis2012}; HD 166191 provides a young, evolving counterpart \citep{Su2022}. HD 69830 and $\eta$ Corvi are useful examples at lower dust luminosities \citep{Lisse2007,Lisse2012}. The range of ages and spectra in recent census and JWST/MIRI work argues against assigning one mechanism from broadband colours alone \citep{Moor2024,Su2026}. For the present source, the interest lies in a substantial warm component around a low-mass host without established youth. A disc and its dust-production mechanism remain to be identified.

\section{Alternative explanations}
The exposure tests make a single-frame artefact an unlikely explanation, and the image models disfavour the known neighbour as the sole emitter. A dusty giant is also difficult to reconcile with the Gaia luminosity and gravity. Primordial material would require supporting evidence for youth or accretion. Warm dust is a plausible explanation for the MIR shape; if it accounts for the full excess, its luminosity calls for an extreme or transient dust population rather than ordinary faint debris.

The main competing association hypothesis is a compact background galaxy or AGN. We test 25 SWIRE templates \citep{Polletta2007}, normalised to the W3 excess, over 41 redshifts from 0 to 2 and five additional extinction values, $A_0=0,1,2,4,8$ mag, using a phenomenological extinction screen \citep{Gordon2023}. In 3853 of the 5125 configurations, the background component alone exceeds the observed total in a supported band by more than three native errors. The remaining 1272 pass this flux-budget test, although passing does not amount to a successful joint SED fit. The rejected 75.18 per cent describes this particular grid, not the probability of AGN contamination. Optical/NIR imaging has no calibrated completeness within the stellar core, leaving a sufficiently close red source viable.

Radio and X-ray data constrain the more readily detectable versions of that scenario. No significant 3-GHz VLASS counterpart is found. Three usable images give three-RMS sensitivities of 0.448, 0.456 and 0.487 mJy, using the larger of local annular scatter and the supplied noise map. Adding positive forced flux gives bounds of 0.448, 0.484 and 0.639 mJy. The eROSITA DR2/eRASS:3 service gives a 0.2--2.3 keV limit of $2.938\times10^{-14}$ erg s$^{-1}$ cm$^{-2}$ for 469.5 s exposure, with a one-sided 99.87 per cent construction and an absorbed power law of $\Gamma=2$, $N_{\rm H}=3\times10^{20}$ cm$^{-2}$. Radio-faint sources can evade the former constraints, and heavy absorption changes the X-ray conversion. Neither these limits nor the incomplete local catalogue counts yield a contamination probability for this object.

The optical record gives a further reason to monitor the field. On 2021 February 25, ZTF measured a faint state approximately 0.180, 0.170 and 0.147 mag below the local $g,r,i$ baselines. Those measurements span 63.6 min, so their relative depths do not establish chromatic dimming. A same-night ASAS-SN point is suggestive, but variable controls and blending weaken its value as independent confirmation. TESS shows 2--4 per cent excursions without a coherent occultation period and cannot resolve the neighbour. The appropriate classification is complex, unresolved variability. Simultaneous resolved monitoring is needed before connecting it to the MIR excess.

\section{Discussion and conclusion}
The W3 excess survives the source-level checks. It appears in every available cryogenic exposure, remains after single-frame removal and is centred near the Gaia star. The known 6.1-arcsec neighbour cannot reproduce the central emission in the tested models. These results make \source\ a strong follow-up target, while leaving one important ambiguity: WISE cannot separate the star from a nearly coincident hidden MIR source.

For a circumstellar interpretation, the inferred temperature and fractional luminosity point to an extreme warm-dust system around a low-mass dwarf. That would be an astrophysically useful result in its own right. No causal natural explanation has yet been demonstrated, but warm dust remains plausible, and the present photometry does not establish its production mechanism. The clearest next observation is subarcsecond imaging at 10--15 $\mu$m, with a measured PSF and accurate registration to Gaia. MIR spectroscopy can then test for dust features, PAH emission or a background redshift. Further MIR epochs would show whether the excess persists, while stellar youth diagnostics would put the dust interpretation on firmer ground.

The source also bears on infrared technosignature searches. The waste-heat argument of \citet{Dyson1960} motivated stellar infrared searches, including the IRAS work of \citet{Carrigan2009}. The G-hat programme developed an energy-budget framework and explored the dusty sources that can mimic these signatures \citep{Wright2014,Wright2014II,Griffith2015}. Project Hephaistos applies this approach to stellar samples and candidate follow-up \citep{Suazo2022,Suazo2024,Korn2026}. A thermal fit alone does not identify the emitter or its energy source. Here, unresolved dust and background-source alternatives prevent an artificial-origin interpretation. Resolving the excess would still be valuable for these searches: it would identify which astrophysical systems survive their selection and how follow-up can distinguish them \citep{Ren2025,Ren2026,Zackrisson2026}.

\section*{Acknowledgements}
This work uses data from ESA's \textit{Gaia} mission \citep{GaiaMission,GaiaDR3}, processed by the Gaia Data Processing and Analysis Consortium (DPAC). National institutions support DPAC through the Gaia Multilateral Agreement. The 2MASS project \citep{Skrutskie2006} is a collaboration between the University of Massachusetts and IPAC/Caltech, funded by NASA and the NSF. WISE \citep{Wright2010} and NEOWISE \citep{Mainzer2014} are NASA-funded projects of UCLA/JPL-Caltech and JPL-Caltech, respectively.

The author thanks the teams providing public ZTF \citep{Bellm2019}, ASAS-SN \citep{Shappee2014,Kochanek2017}, TESS \citep{Ricker2015} and VLASS \citep{Lacy2020} data. NASA's Science Mission Directorate funds TESS. NRAO is an NSF facility operated by Associated Universities, Inc. This work also uses eROSITA-DE products from SRG/eROSITA \citep{Predehl2021,RamosCeja2026} and its upper-limit service \citep{Tubin2024}. Catalogue access used CDS VizieR, Strasbourg \citep{Ochsenbein2000}, and NASA/IPAC IRSA, operated by Caltech under contract with NASA. TESS data were obtained through MAST at STScI.

OpenAI tools were used for code-development assistance and language editing. The author takes full responsibility for the final manuscript.

\section*{Data availability}
Catalogue identifiers, source values, per-exposure measurements and model outputs will be made available. The input products are available through the \href{https://gea.esac.esa.int/archive/}{Gaia Archive}, \href{https://irsa.ipac.caltech.edu/}{NASA/IPAC IRSA}, \href{https://asas-sn.osu.edu/}{ASAS-SN}, \href{https://mast.stsci.edu/}{MAST}, the \href{https://science.nrao.edu/vlass/data-access}{VLASS archive}, \href{https://erosita.mpe.mpg.de/erodat/}{eRODat} and \href{https://vizier.cds.unistra.fr/}{CDS VizieR}. The eROSITA DR2 products used here comprise catalogue information and upper limits \citep{RamosCeja2026,Tubin2024}. The accompanying provenance records the input products and analysis scripts. The analysis products will be released with the forthcoming open-source Dyson Atlas infrared technosignature survey.

\bibliographystyle{mnras}
\bibliography{references}

@article{Moor2021,
  author = {Mo{\'{o}}r, Attila and {\'{A}}brah{\'{a}}m, P{\'{e}}ter and Szab{\'{o}}, Gyula and Vida, Kriszti{\'{a}}n and Cataldi, Gianni and Derekas, Al{\'{i}}z and Henning, Thomas and Kinemuchi, Karen and others},
  year = {2021},
  title = {{A New Sample of Warm Extreme Debris Disks from the ALLWISE Catalog}},
  journal = {ApJ},
  volume = {910},
  pages = {27},
  doi = {10.3847/1538-4357/abdc26},
  url = {https://doi.org/10.3847/1538-4357/abdc26},
  eprint = {2103.00568},
  archivePrefix = {arXiv}
}

@article{Moor2024,
  author = {Mo{\'{o}}r, Attila and {\'{A}}brah{\'{a}}m, P{\'{e}}ter and Su, Kate Y L and Henning, Thomas and Marino, Sebastian and Chen, Lei and K{\'{o}}sp{\'{a}}l, {\'{A}}gnes and Pawellek, Nicole and others},
  year = {2024},
  title = {{Abundant sub-micron grains revealed in newly discovered extreme debris discs}},
  journal = {MNRAS},
  volume = {528},
  pages = {4528--4546},
  doi = {10.1093/mnras/stae155},
  url = {https://doi.org/10.1093/mnras/stae155},
  eprint = {2402.15438},
  archivePrefix = {arXiv}
}

@article{KennedyWyatt2013,
  author = {Kennedy, G. M. and Wyatt, M. C.},
  year = {2013},
  title = {{The bright end of the exo-Zodi luminosity function: disc evolution and implications for exo-Earth detectability}},
  journal = {MNRAS},
  volume = {433},
  pages = {2334--2356},
  doi = {10.1093/mnras/stt900},
  url = {https://doi.org/10.1093/mnras/stt900},
  eprint = {1305.6607},
  archivePrefix = {arXiv}
}

@article{Avenhaus2012,
  author = {Avenhaus, H. and Schmid, H. M. and Meyer, M. R.},
  year = {2012},
  title = {{The nearby population of M-dwarfs with WISE: a search for warm circumstellar dust}},
  journal = {A\&A},
  volume = {548},
  pages = {A105},
  doi = {10.1051/0004-6361/201219783},
  url = {https://doi.org/10.1051/0004-6361/201219783},
  eprint = {1209.0678},
  archivePrefix = {arXiv}
}

@article{BinksJeffries2017,
  author = {Binks, A. S. and Jeffries, R. D.},
  year = {2017},
  title = {{A WISE-based search for debris discs amongst M dwarfs in nearby, young, moving groups}},
  journal = {MNRAS},
  volume = {469},
  pages = {579--593},
  doi = {10.1093/mnras/stx838},
  url = {https://doi.org/10.1093/mnras/stx838},
  eprint = {1611.07416},
  archivePrefix = {arXiv}
}

@article{Song2005,
  author = {Song, Inseok and Zuckerman, B. and Weinberger, Alycia J. and Becklin, E. E.},
  year = {2005},
  title = {{Extreme collisions between planetesimals as the origin of warm dust around a Sun-like star}},
  journal = {Nature},
  volume = {436},
  pages = {363--365},
  doi = {10.1038/nature03853},
  url = {https://doi.org/10.1038/nature03853}
}

@article{Melis2021,
  author = {Melis, Carl and Olofsson, Johan and Song, Inseok and Sarkis, Paula and Weinberger, Alycia J. and Kennedy, Grant and Krumpe, Mirko},
  year = {2021},
  title = {{Highly Structured Inner Planetary System Debris around the Intermediate Age Sun-like Star TYC 8830 410 1}},
  journal = {ApJ},
  volume = {923},
  pages = {90},
  doi = {10.3847/1538-4357/ac2603},
  url = {https://doi.org/10.3847/1538-4357/ac2603},
  eprint = {2104.06448},
  archivePrefix = {arXiv}
}

@article{Meng2015,
  author = {Meng, Huan Y. A. and Su, Kate Y. L. and Rieke, George H. and Rujopakarn, Wiphu and Myers, Gordon and Cook, Michael and Erdelyi, Emery and Maloney, Chris and others},
  year = {2015},
  title = {{PLANETARY COLLISIONS OUTSIDE THE SOLAR SYSTEM: TIME DOMAIN CHARACTERIZATION OF EXTREME DEBRIS DISKS}},
  journal = {ApJ},
  volume = {805},
  pages = {77},
  doi = {10.1088/0004-637X/805/1/77},
  url = {https://doi.org/10.1088/0004-637X/805/1/77},
  eprint = {1503.05610},
  archivePrefix = {arXiv}
}

@article{Rieke2021,
  author = {Rieke, G. H. and Su, K. Y. L. and Melis, Carl and G{\'{a}}sp{\'{a}}r, Andr{\'{a}}s},
  year = {2021},
  title = {{Extreme Variability of the V488 Persei Debris Disk}},
  journal = {ApJ},
  volume = {918},
  pages = {71},
  doi = {10.3847/1538-4357/ac0dc4},
  url = {https://doi.org/10.3847/1538-4357/ac0dc4},
  eprint = {2108.02901},
  archivePrefix = {arXiv}
}

@article{deWit2013,
  author = {\VAN{Wit}{de}{de} Wit, W. J. and Grinin, V. P. and Potravnov, I. S. and Shakhovskoi, D. N. and M{\"{u}}ller, A. and Moerchen, M.},
  year = {2013},
  title = {{Active asteroid belt causes the UXOR phenomenon in RZ Piscium}},
  journal = {A\&A},
  volume = {553},
  pages = {L1},
  doi = {10.1051/0004-6361/201220715},
  url = {https://doi.org/10.1051/0004-6361/201220715},
  eprint = {1303.4138},
  archivePrefix = {arXiv}
}

@article{Melis2012,
  author = {Melis, Carl and Zuckerman, B. and Rhee, Joseph H. and Song, Inseok and Murphy, Simon J. and Bessell, Michael S.},
  year = {2012},
  title = {{Rapid disappearance of a warm, dusty circumstellar disk}},
  journal = {Nature},
  volume = {487},
  pages = {74--76},
  doi = {10.1038/nature11210},
  url = {https://doi.org/10.1038/nature11210},
  eprint = {1207.1162},
  archivePrefix = {arXiv}
}

@article{Lisse2007,
  author = {Lisse, C. M. and Beichman, C. A. and Bryden, G. and Wyatt, M. C.},
  year = {2007},
  title = {{On the Nature of the Dust in the Debris Disk around HD 69830}},
  journal = {ApJ},
  volume = {658},
  pages = {584--592},
  doi = {10.1086/511001},
  url = {https://doi.org/10.1086/511001},
  eprint = {astro-ph/0611452},
  archivePrefix = {arXiv}
}

@article{Lisse2012,
  author = {Lisse, C. M. and Wyatt, M. C. and Chen, C. H. and Morlok, A. and Watson, D. M. and Manoj, P. and Sheehan, P. and Currie, T. M. and others},
  year = {2012},
  title = {{SPITZER EVIDENCE FOR A LATE-HEAVY BOMBARDMENT AND THE FORMATION OF UREILITES IN $\eta$ CORVI At $\sim$1 Gyr}},
  journal = {ApJ},
  volume = {747},
  pages = {93},
  doi = {10.1088/0004-637X/747/2/93},
  url = {https://doi.org/10.1088/0004-637X/747/2/93},
  eprint = {1110.4172},
  archivePrefix = {arXiv}
}

@article{Su2022,
  author = {Su, Kate Y. L. and Kennedy, Grant M. and Schlawin, Everett and Jackson, Alan P. and Rieke, G. H.},
  year = {2022},
  title = {{A Star-sized Impact-produced Dust Clump in the Terrestrial Zone of the HD 166191 System}},
  journal = {ApJ},
  volume = {927},
  pages = {135},
  doi = {10.3847/1538-4357/ac4bbb},
  url = {https://doi.org/10.3847/1538-4357/ac4bbb},
  eprint = {2203.02366},
  archivePrefix = {arXiv}
}

@article{Wright2014,
  author = {Wright, J. T. and Mullan, B. and Sigurdsson, S. and Povich, M. S.},
  year = {2014},
  title = {{THE $\hat{G}$ INFRARED SEARCH FOR EXTRATERRESTRIAL CIVILIZATIONS WITH LARGE ENERGY SUPPLIES. I. BACKGROUND AND JUSTIFICATION}},
  journal = {ApJ},
  volume = {792},
  pages = {26},
  doi = {10.1088/0004-637X/792/1/26},
  url = {https://doi.org/10.1088/0004-637X/792/1/26},
  eprint = {1408.1133},
  archivePrefix = {arXiv}
}

@article{Suazo2022,
  author = {Suazo, Mat{\'{i}}as and Zackrisson, Erik and Wright, Jason T and Korn, Andreas J and Huston, Macy},
  year = {2022},
  title = {{Project Hephaistos -- I. Upper limits on partial Dyson spheres in the Milky Way}},
  journal = {MNRAS},
  volume = {512},
  pages = {2988--3000},
  doi = {10.1093/mnras/stac280},
  url = {https://doi.org/10.1093/mnras/stac280},
  eprint = {2201.11123},
  archivePrefix = {arXiv}
}

@article{Suazo2024,
  author = {Suazo, Mat{\'{i}}as and Zackrisson, Erik and Mahto, Priyatam K and Lundell, Fabian and Nettelblad, Carl and Korn, Andreas J and Wright, Jason T and Majumdar, Suman},
  year = {2024},
  title = {{Project Hephaistos -- II. Dyson sphere candidates from Gaia DR3, 2MASS, and WISE}},
  journal = {MNRAS},
  volume = {531},
  pages = {695--707},
  doi = {10.1093/mnras/stae1186},
  url = {https://doi.org/10.1093/mnras/stae1186},
  eprint = {2405.02927},
  archivePrefix = {arXiv}
}

@article{GaiaDR3,
  author = {{Gaia Collaboration} and others},
  year = {2023},
  title = {{Gaia Data Release 3: Summary of the content and survey properties}},
  journal = {A\&A},
  volume = {674},
  pages = {A1},
  doi = {10.1051/0004-6361/202243940},
  url = {https://doi.org/10.1051/0004-6361/202243940},
  eprint = {2208.00211},
  archivePrefix = {arXiv}
}

@article{Creevey2023,
  author = {Creevey, O. L. and Sordo, R. and Pailler, F. and Fr{\'{e}}mat, Y. and Heiter, U. and Th{\'{e}}venin, F. and Andrae, R. and Fouesneau, M. and others},
  year = {2023},
  title = {{Gaia Data Release 3: Astrophysical parameters inference system (Apsis). I. Methods and content overview}},
  journal = {A\&A},
  volume = {674},
  pages = {A26},
  doi = {10.1051/0004-6361/202243688},
  url = {https://doi.org/10.1051/0004-6361/202243688}
}

@article{Skrutskie2006,
  author = {Skrutskie, M. F. and Cutri, R. M. and Stiening, R. and Weinberg, M. D. and Schneider, S. and Carpenter, J. M. and Beichman, C. and Capps, R. and others},
  year = {2006},
  title = {{The Two Micron All Sky Survey (2MASS)}},
  journal = {AJ},
  volume = {131},
  pages = {1163--1183},
  doi = {10.1086/498708},
  url = {https://doi.org/10.1086/498708}
}

@article{Wright2010,
  author = {Wright, Edward L. and Eisenhardt, Peter R. M. and Mainzer, Amy K. and Ressler, Michael E. and Cutri, Roc M. and Jarrett, Thomas and Kirkpatrick, J. Davy and Padgett, Deborah and others},
  year = {2010},
  title = {{THE WIDE-FIELD INFRARED SURVEY EXPLORER (WISE): MISSION DESCRIPTION AND INITIAL ON-ORBIT PERFORMANCE}},
  journal = {AJ},
  volume = {140},
  pages = {1868--1881},
  doi = {10.1088/0004-6256/140/6/1868},
  url = {https://doi.org/10.1088/0004-6256/140/6/1868},
  eprint = {1008.0031},
  archivePrefix = {arXiv}
}

@article{Meszaros2024,
  author = {M{\'{e}}sz{\'{a}}ros, Szabolcs and Bohlin, Ralph and Allende Prieto, Carlos and Cseh, Borb{\'{a}}la and Kov{\'{a}}cs, J{\'{o}}zsef and Fleming, Scott W. and Dencs, Zolt{\'{a}}n and Deustua, Susana and others},
  year = {2024},
  title = {{The updated BOSZ synthetic stellar spectral library}},
  journal = {A\&A},
  volume = {688},
  pages = {A197},
  doi = {10.1051/0004-6361/202449306},
  url = {https://doi.org/10.1051/0004-6361/202449306},
  eprint = {2407.10872},
  archivePrefix = {arXiv}
}

@article{Polletta2007,
  author = {Polletta, M. and Tajer, M. and Maraschi, L. and Trinchieri, G. and Lonsdale, C. J. and Chiappetti, L. and Andreon, S. and Pierre, M. and others},
  year = {2007},
  title = {{Spectral Energy Distributions of Hard X-Ray Selected Active Galactic Nuclei in the XMM-Newton Medium Deep Survey}},
  journal = {ApJ},
  volume = {663},
  pages = {81--102},
  doi = {10.1086/518113},
  url = {https://doi.org/10.1086/518113},
  eprint = {astro-ph/0703255},
  archivePrefix = {arXiv}
}

@article{Gordon2023,
  author = {Gordon, Karl D. and Clayton, Geoffrey C. and Decleir, Marjorie and Fitzpatrick, E. L. and Massa, Derck and Misselt, Karl A. and Tollerud, Erik J.},
  year = {2023},
  title = {{One Relation for All Wavelengths: The Far-ultraviolet to Mid-infrared Milky Way Spectroscopic R(V)-dependent Dust Extinction Relationship}},
  journal = {ApJ},
  volume = {950},
  pages = {86},
  doi = {10.3847/1538-4357/accb59},
  url = {https://doi.org/10.3847/1538-4357/accb59},
  eprint = {2304.01991},
  archivePrefix = {arXiv}
}

@article{Su2026,
  author = {Su, Kate Y. L. and Moor, Attila and Kospal, Agnes and Rieke, George H. and Sefilian, Antranik A. and Malhotra, Renu and Pascucci, Ilaria and Jackson, Alan P. and others},
  year = {2026},
  title = {{Extreme Debris Disks: Insights into Violent Collisions in Planet Formation and Destruction}},
  journal = {arXiv e-prints},
  pages = {arXiv:2607.06684},
  eprint = {2607.06684},
  archivePrefix = {arXiv},
  doi = {10.48550/arXiv.2607.06684},
  url = {https://arxiv.org/abs/2607.06684},
  note = {accepted for publication in ApJ}
}

@article{Korn2026,
  author = {Korn, Andreas J. and Suazo, Mat{\'{i}}as and Zackrisson, Erik and Cort{\'{e}}s-Zuleta, P{\'{i}}a and Rains, Adam D. and Nabizadeh, Armin},
  year = {2026},
  title = {{Project Hephaistos -- III. Characterizing anomalous infrared sources identified as Dyson-sphere candidates}},
  journal = {arXiv e-prints},
  pages = {arXiv:2607.25701},
  eprint = {2607.25701},
  archivePrefix = {arXiv},
  doi = {10.48550/arXiv.2607.25701},
  url = {https://arxiv.org/abs/2607.25701},
  note = {arXiv preprint}
}

@article{Zackrisson2026,
  author = {Zackrisson, Erik and Bik, Arjan and Asgar, Anita Ali and Curtis, Olivia and Wright, Jason T. and Ren, Tongtian and Assef, Roberto J. and Blain, Andrew and others},
  year = {2026},
  title = {{Project Hephaistos -- IV. James Webb Space Telescope Observations of Two Dyson Sphere Candidates}},
  journal = {arXiv e-prints},
  pages = {arXiv:2607.09460},
  eprint = {2607.09460},
  archivePrefix = {arXiv},
  doi = {10.48550/arXiv.2607.09460},
  url = {https://arxiv.org/abs/2607.09460},
  note = {arXiv preprint}
}

@article{Ren2026,
  author = {Ren, Tongtian and Garrett, Michael A. and Zackrisson, Erik and Korn, Andreas J. and Siemion, Andrew P. V. and Wright, Jason T. and Brandeker, Alexis},
  year = {2026},
  title = {{Archival Diagnostics for Potential Background Contaminantsof Project Hephaistos Dyson Sphere Candidates}},
  journal = {arXiv e-prints},
  pages = {arXiv:2607.03619},
  eprint = {2607.03619},
  archivePrefix = {arXiv},
  doi = {10.48550/arXiv.2607.03619},
  url = {https://arxiv.org/abs/2607.03619},
  note = {arXiv preprint, version 5}
}

@article{Gagne2018,
  author = {Gagn{\'e}, J. and Mamajek, E. E. and Malo, L. and others},
  title = {{BANYAN. XI. The BANYAN $\Sigma$ Multivariate Bayesian Algorithm to Identify Members of Young Associations within 150 pc}},
  journal = {ApJ},
  year = {2018},
  volume = {856},
  pages = {23},
  doi = {10.3847/1538-4357/aaae09},
  eprint = {1801.09051},
  archivePrefix = {arXiv}
}

@article{Gagne2026,
  author = {Gagn{\'e}, J. and Moranta, L. and Faherty, J. K. and others},
  title = {{The Montreal Open Clusters and Associations (MOCA) Database: A Census of Nearby Associations, Open Clusters, and Young Substellar Objects within 500 pc of the Sun}},
  journal = {arXiv e-prints},
  year = {2026},
  pages = {arXiv:2602.15695},
  eprint = {2602.15695},
  archivePrefix = {arXiv},
  doi = {10.48550/arXiv.2602.15695},
  note = {accepted for publication in ApJS}
}

@article{Kerr2023,
  author = {Kerr, R. and Kraus, A. L. and Rizzuto, A. C.},
  title = {{SPYGLASS. IV. New Stellar Survey of Recent Star Formation within 1 kpc}},
  journal = {ApJ},
  year = {2023},
  volume = {954},
  pages = {134},
  doi = {10.3847/1538-4357/ace5b3},
  eprint = {2306.08150},
  archivePrefix = {arXiv}
}

@ARTICLE{Usatov2018,
       author = {{Usatov}, Maxim},
        title = "{Optically Variable Quasars in the AllWISE and Pan-STARRS1 Data}",
      journal = {Journal of Astronomical Data},
         year = 2018,
        month = dec,
       volume = {24},
        pages = {3},
       adsurl = {https://ui.adsabs.harvard.edu/abs/2018JAD....24....3U}
}

@article{GaiaMission,
  author = {{Gaia Collaboration} and others},
  year = {2016},
  title = {{The Gaia mission}},
  journal = {A\&A},
  volume = {595},
  pages = {A1},
  doi = {10.1051/0004-6361/201629272},
  url = {https://doi.org/10.1051/0004-6361/201629272},
  eprint = {1609.04153},
  archivePrefix = {arXiv}
}

@article{Mainzer2014,
  author = {Mainzer, A. and others},
  year = {2014},
  title = {{Initial Performance of the NEOWISE Reactivation Mission}},
  journal = {ApJ},
  volume = {792},
  pages = {30},
  doi = {10.1088/0004-637X/792/1/30},
  url = {https://doi.org/10.1088/0004-637X/792/1/30},
  eprint = {1406.6025},
  archivePrefix = {arXiv}
}

@article{Bellm2019,
  author = {Bellm, E. C. and others},
  year = {2019},
  title = {{The Zwicky Transient Facility: System Overview, Performance, and First Results}},
  journal = {PASP},
  volume = {131},
  pages = {018002},
  doi = {10.1088/1538-3873/aaecbe},
  url = {https://doi.org/10.1088/1538-3873/aaecbe}
}

@article{Shappee2014,
  author = {Shappee, B. J. and others},
  year = {2014},
  title = {{The Man Behind the Curtain: X-rays Drive the UV through NIR Variability in the 2013 AGN Outburst in NGC 2617}},
  journal = {ApJ},
  volume = {788},
  pages = {48},
  doi = {10.1088/0004-637X/788/1/48},
  url = {https://doi.org/10.1088/0004-637X/788/1/48},
  eprint = {1310.2241},
  archivePrefix = {arXiv}
}

@article{Kochanek2017,
  author = {Kochanek, C. S. and others},
  year = {2017},
  title = {{The All-Sky Automated Survey for Supernovae (ASAS-SN) Light Curve Server v1.0}},
  journal = {PASP},
  volume = {129},
  pages = {104502},
  doi = {10.1088/1538-3873/aa80d9},
  url = {https://doi.org/10.1088/1538-3873/aa80d9},
  eprint = {1706.07060},
  archivePrefix = {arXiv}
}

@article{Ricker2015,
  author = {Ricker, G. R. and others},
  year = {2015},
  title = {{Transiting Exoplanet Survey Satellite}},
  journal = {J. Astron. Telesc. Instrum. Syst.},
  volume = {1},
  pages = {014003},
  doi = {10.1117/1.JATIS.1.1.014003},
  url = {https://doi.org/10.1117/1.JATIS.1.1.014003},
  eprint = {1406.0151},
  archivePrefix = {arXiv}
}

@article{Lacy2020,
  author = {Lacy, M. and others},
  year = {2020},
  title = {{The Karl G. Jansky Very Large Array Sky Survey (VLASS). Science Case and Survey Design}},
  journal = {PASP},
  volume = {132},
  pages = {035001},
  doi = {10.1088/1538-3873/ab63eb},
  url = {https://doi.org/10.1088/1538-3873/ab63eb},
  eprint = {1907.01981},
  archivePrefix = {arXiv}
}

@article{Predehl2021,
  author = {Predehl, P. and others},
  year = {2021},
  title = {{The eROSITA X-ray telescope on SRG}},
  journal = {A\&A},
  volume = {647},
  pages = {A1},
  doi = {10.1051/0004-6361/202039313},
  url = {https://doi.org/10.1051/0004-6361/202039313},
  eprint = {2010.03477},
  archivePrefix = {arXiv}
}

@article{Tubin2024,
  author = {Tub{\'i}n-Arenas, D. and others},
  year = {2024},
  title = {{The eROSITA upper limits: Description and access to the data}},
  journal = {A\&A},
  volume = {682},
  pages = {A35},
  doi = {10.1051/0004-6361/202346773},
  url = {https://doi.org/10.1051/0004-6361/202346773},
  eprint = {2401.17305},
  archivePrefix = {arXiv}
}

@article{RamosCeja2026,
  author = {Ramos-Ceja, M. E. and others},
  year = {2026},
  title = {{The SRG/eROSITA All-Sky Survey DR2: Cumulative X-ray catalogues from the first three surveys and multi-wavelength counterparts in the western Galactic hemisphere}},
  journal = {arXiv e-prints},
  pages = {arXiv:2607.27772},
  doi = {10.48550/arXiv.2607.27772},
  url = {https://doi.org/10.48550/arXiv.2607.27772},
  eprint = {2607.27772},
  archivePrefix = {arXiv},
  note = {accepted for publication in A\&A}
}

@article{Ochsenbein2000,
  author = {Ochsenbein, F. and Bauer, P. and Marcout, J.},
  year = {2000},
  title = {{The VizieR database of astronomical catalogues}},
  journal = {A\&AS},
  volume = {143},
  pages = {23--32},
  doi = {10.1051/aas:2000169},
  url = {https://doi.org/10.1051/aas:2000169}
}

@article{Dyson1960,
  author = {Dyson, F. J.},
  year = {1960},
  title = {{Search for Artificial Stellar Sources of Infrared Radiation}},
  journal = {Science},
  volume = {131},
  pages = {1667--1668},
  doi = {10.1126/science.131.3414.1667},
  url = {https://pubmed.ncbi.nlm.nih.gov/17780673/}
}

@article{Carrigan2009,
  author = {Carrigan, Jr., R. A.},
  year = {2009},
  title = {{IRAS-based Whole-Sky Upper Limit on Dyson Spheres}},
  journal = {ApJ},
  volume = {698},
  pages = {2075--2086},
  doi = {10.1088/0004-637X/698/2/2075},
  url = {https://arxiv.org/abs/0811.2376},
  eprint = {0811.2376},
  archivePrefix = {arXiv}
}

@article{Wright2014II,
  author = {Wright, J. T. and Griffith, R. L. and Sigurdsson, S. and Povich, M. S. and Mullan, B.},
  year = {2014},
  title = {{The $\hat{G}$ Infrared Search for Extraterrestrial Civilizations with Large Energy Supplies. II. Framework, Strategy, and First Result}},
  journal = {ApJ},
  volume = {792},
  pages = {27},
  doi = {10.1088/0004-637X/792/1/27},
  url = {https://arxiv.org/abs/1408.1134},
  eprint = {1408.1134},
  archivePrefix = {arXiv}
}

@article{Griffith2015,
  author = {Griffith, R. L. and Wright, J. T. and Maldonado, J. and Povich, M. S. and Sigurdsson, S. and Mullan, B.},
  year = {2015},
  title = {{The $\hat{G}$ Infrared Search for Extraterrestrial Civilizations with Large Energy Supplies. III. The Reddest Extended Sources in WISE}},
  journal = {ApJS},
  volume = {217},
  pages = {25},
  doi = {10.1088/0067-0049/217/2/25},
  url = {https://arxiv.org/abs/1504.03418},
  eprint = {1504.03418},
  archivePrefix = {arXiv}
}

@article{Contardo2024,
  author = {Contardo, G. and Hogg, D. W.},
  year = {2024},
  title = {{A Data-driven Search For Mid-infrared Excesses Among Five Million Main-sequence FGK Stars}},
  journal = {AJ},
  volume = {168},
  pages = {157},
  doi = {10.3847/1538-3881/ad6b90},
  url = {https://arxiv.org/abs/2403.18941},
  eprint = {2403.18941},
  archivePrefix = {arXiv}
}

@article{Ren2024,
  author = {Ren, T. and Garrett, M. A. and Siemion, A. P. V.},
  year = {2024},
  title = {{Background Contamination of the Project Hephaistos Dyson Spheres Candidates}},
  journal = {arXiv e-prints},
  pages = {arXiv:2405.14921},
  doi = {10.48550/arXiv.2405.14921},
  url = {https://arxiv.org/abs/2405.14921},
  eprint = {2405.14921},
  archivePrefix = {arXiv}
}

@article{Ren2025,
  author = {Ren, T. and Garrett, M. A. and Siemion, A. P. V.},
  year = {2025},
  title = {{High-resolution imaging of the radio source associated with Project Hephaistos Dyson Sphere Candidate G}},
  journal = {MNRAS},
  volume = {538},
  pages = {L56--L61},
  doi = {10.1093/mnrasl/slaf006},
  url = {https://arxiv.org/abs/2501.05152},
  eprint = {2501.05152},
  archivePrefix = {arXiv}
}

@article{Blain2024,
  author = {Blain, A. W.},
  year = {2024},
  title = {{Did WISE detect Dyson Spheres/Structures around Gaia-2MASS-selected stars?}},
  journal = {arXiv e-prints},
  pages = {arXiv:2409.11447},
  doi = {10.48550/arXiv.2409.11447},
  url = {https://arxiv.org/abs/2409.11447},
  eprint = {2409.11447},
  archivePrefix = {arXiv}
}
\bsp
\label{lastpage}
\end{document}